\documentclass[preprint2]{aastex} 
\usepackage{amssymb, amsmath, graphics} 
\usepackage{longtable,graphicx} 
\usepackage{aas_macros} 
 
\usepackage[usenames]{color}

\shorttitle{High-degree prograde modes} 
\shortauthors{Michel Breger et al.}
 
\begin{document}

\title{Detection of high-degree prograde sectoral mode sequences in the A-star KIC 8054146?}

\author
{M.~Breger\altaffilmark{1,2}, P. Lenz\altaffilmark{3,2}, A. A. Pamyatnykh\altaffilmark{3}}
\altaffiltext{1}{Department of Astronomy, University of Texas, Austin, TX 78712, USA}
\altaffiltext{2}{Institut f\"ur Astronomie der Universit\"at Wien, T\"urkenschanzstr. 17, A--1180, Wien, Austria}
\altaffiltext{3}{Copernicus Astronomical Center, Bartycka 18, 00-716 Warsaw, Poland}

\begin{abstract} 

This paper examines the 46 frequencies found in the $\delta$ Sct star KIC 8054146 involving a frequency spacing
of $\it{exactly}$ 2.814 cycles day$^{-1}$ (32.57 $\mu$Hz), which is also a dominant
low-frequency peak near or equal to the rotational frequency. These 46 frequencies range up to 146 cycles day$^{-1}$.
Three years of {\it Kepler} data reveal distinct sequences of these equidistantly spaced frequencies, including the basic sequence and
side lobes associated with other dominant modes (i.e., small amplitude modulations).
The amplitudes of the basic sequence show a high-low pattern. The basic sequence follows the equation
$f_m = 2.8519 + m * 2.81421$ cycles day$^{-1}$ with $m$ ranging from 25 to 35. The zero-point offset and the
lack of low-order harmonics eliminate an interpretation in terms of a Fourier series of a non-sinusoidal light curve.
The exactness of the spacing eliminates high-order asymptotic pulsation. The frequency pattern is not compatible
with simple hypotheses involving single or multiple spots, even with differential rotation.

The basic high-frequency sequence is interpreted in terms of prograde sectoral modes. These can be marginally unstable,
while their corresponding low-degree counterparts are stable due to stronger damping. The measured projected rotation
velocity (300 km s$^{-1}$) indicates that the star rotates with $\gtrapprox$ 70\% of the Keplerian
break-up velocity. This suggests a near equator-on view. We qualitatively examine the visibility of prograde sectoral high-degree
g-modes in integrated photometric light in such a geometrical configuration and find that prograde sectoral modes
can reproduce the frequencies and the odd-even amplitude pattern of the high-frequency sequence.\end{abstract} 
 

\section{Introduction} 
 
\begin{figure*}[htbp]
\centering 
\includegraphics[bb=17 17 770 538,clip,width=\textwidth]{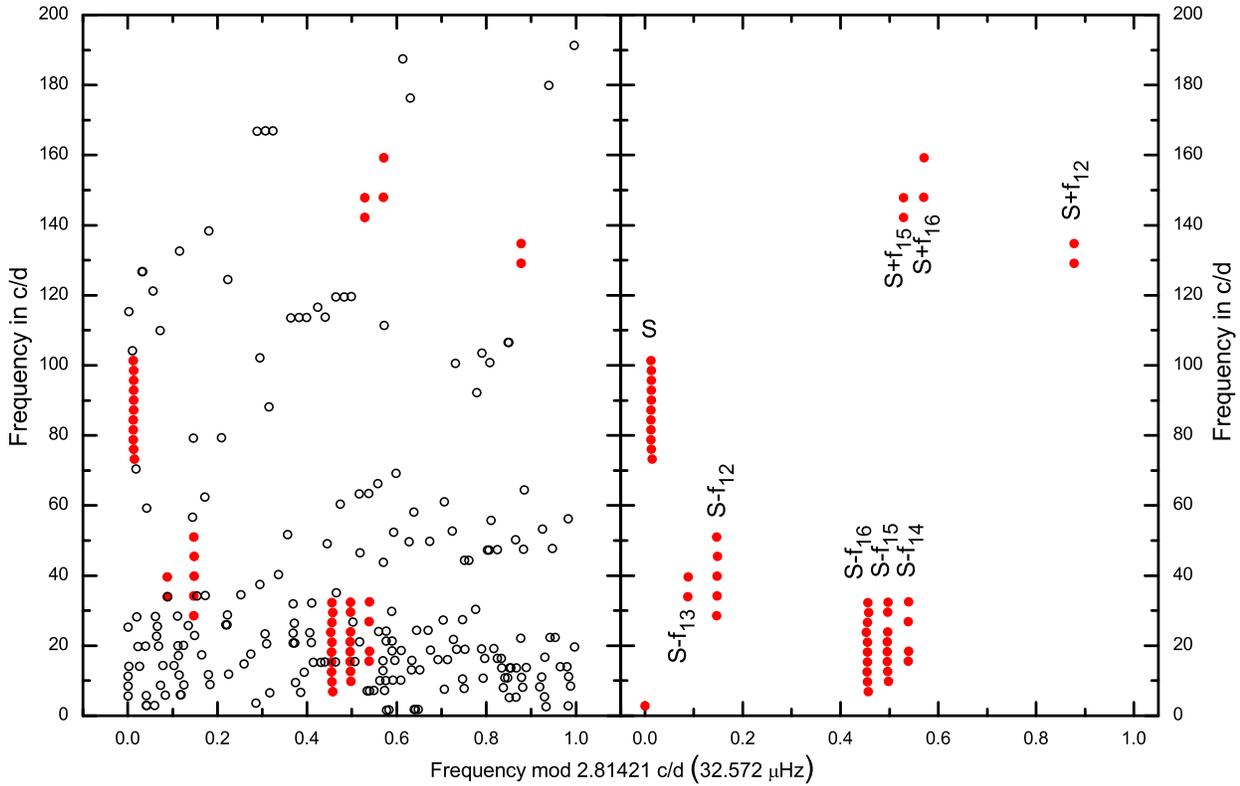} 
\caption{Echelle diagram of the detected frequencies of KIC 8054146 folded with 2.81421 cycles day$^{-1}$ (left panel). The filled circles are the frequency peaks, which are identified as part of the S sequences through the equidistant spacings as well as their actual frequency values. Nine separate sequences, S, can be identified. The identifications  of the sequences are given in the right panel,
 e.g., $S-f_{16}$ denotes the S sequence seen as side lobes of the dominant pulsation frequency $f_{16}$. This can be interpreted as the basic sequence, S, and modulation of the amplitudes of other pulsation modes, seen in the frequency diagram as side lobes on both sides of the parent frequencies.}\label{fig:echelle}
\end{figure*} 

Since almost all stars are observed as pinpoints of light, measurements of stellar variability correspond to the light from the integrated stellar disk. Consequently, changes due to fine patterns experience a severely reduced amplitude and may no longer be measurable. Photometrically, the nonradial pulsation modes of high-degree ($\ell\geq$ 3) have been considered to be almost unobservable from the ground due to cancellation effects. This simplistic view was challenged by \citet{jdd2006}, who argued that some of the multiple frequencies with small amplitudes detected during the multisite campaign of the $\delta$ Sct FG Vir \citep{B05} could be high-degree modes. With the advent of photometric measurements from space and the resulting detection of amplitudes of 1 part-per-million (ppm), high-degree modes need to be considered.

Spectroscopically, high-degree modes can be detected through the changes of line profiles, e.g., \cite{Z2006}. The method requires large amounts of observing time on medium-size telescopes. It is presently applied to a number of $\delta$ Sct stars (HR 1170, 4 CVn and EE Cam), for which a number of high-degree modes are detected. It is remarkable that with considerably fewer data, high-degree modes in $\delta$ Sct stars have been seen before: the {\it Musicos} multisite project \citep{Kennelly96} presented evidence for the presence of high-degree modes in a number of $\delta$ Sct stars such as $\tau$ Peg \citep{Kennelly98}. They proposed explanations in terms of prograde sectoral modes of $\ell$ = $\mid$$m$$\mid$ up to 20. These modes would have frequencies that are about equal in the co-rotating frame of the star, leading to equidistant spacing in the observer's frame of reference. From photometry, prograde sectoral modes have recently also been suggested for the star Rasalhague \citep{monnier2010}, which is also a rapidly rotating A star seen equator-on. The authors report $\mid$$m$$\mid$ up to 7.

In this paper we examine whether such modes can be seen in the extensive photometry of the rapidly rotating $\delta$ Sct star KIC 8054146 and relate the observational results to theoretical models. Both the observational and theoretical aspects are pioneering. In the modeling of this star, one has to consider that rotation has a strong influence on the stellar structure. The shape of a rapidly rotating star resembles that of an oblate spheroid rather than a sphere due to the effect of centrifugal force. In fact, the flattening of some rapidly rotating stars could already be empirically measured by means of long-baseline interferometers \citep[for a review see][]{vanBelle2012}.
The oblateness of the stellar surface leads to a latitude dependence of effective gravity and effective temperature. Taken together, these effects are termed gravity darkening \citep{espinosa2011}. 
This leads to the problem that the observed fundamental parameters in rapidly rotating stars, such as effective temperature, gravity and luminosity are a function of aspect, where the angle $i$ between the line-of-sight and the stellar rotation axis is often unknown. The effects of the stellar inclination angle on the effective temperature and luminosity of rotating stellar models were examined, e.g., by \cite{gillich2008}. They found a strong dependence of the fundamental parameters on the aspect. Moreover, it was found by \cite{townsend2004} that for rapid rotation the line-widths become insensitive to rotation, and hence the projected rotational velocity $v\sin i$ is underestimated. While the authors found these results for rapidly rotating Be stars, the same problem may also apply to A stars.

Rotation also has an impact on pulsation. The oscillation cavities are affected by the changes in stellar structure (i.e., the deformation of the star and the decreased mean density). Moreover, the impact of centrifugal and Coriolis force on the oscillations increases with the rotation rate. Rotation, in particular, affects acoustic modes propagating in the envelope \citep{lignieres2009}. 

In the interpretation of observed frequencies in rapidly rotating stars another problem arises because of the effect of rotational splitting \citep[e.g.,][]{aap2003}. 
Due to the transformation from the co-rotating system (star's system) to the observer's system, the mode frequencies are shifted by $mf_{\rm rot}$. For example, for an A type star with an equatorial rotation velocity of 300 km s$^{-1}$, the angular rotation frequency $f_{\rm rot}$ ranges around 2.8 cycles day$^{-1}$. Usually, $m$ is not known a priori and, therefore, a large uncertainty about the intrinsic mode frequency (and its nature) in the co-rotating system of the star persists. In specific cases, however, hints on $m$ may be derived from regular patterns. The rapidly rotating A-star KIC 8054146 with its intriguing frequency pattern  represents such a case.

\section{Results of the {\it Kepler} measurements of KIC 8054146}

The $\delta$ Sct star KIC 8054146 was studied extensively with the {\it Kepler} spacecraft. 7.3 quarters of short-cadence data (one measurement per 58.8s) and 11 quarters of long-cadence (two
measurements per hour) are available. The data were analyzed by \cite{breger2012}. The 349 statistically significant frequencies have amplitudes as low as 2 ppm. One of the most striking results is that several regularities are present in the values of the frequencies. We can rule out the possibility that these regularities are accidental agreements because of the high frequency resolution and large number of exact agreements, as shown in Figure 4 of \cite{breger2012}.

In particular, the frequency patterns seen in the pressure-mode frequency regions (frequencies in the 6 to 200 cycles day$^{-1}$ range) are related to
the dominant low-frequency peaks, interpreted as rotation and gravity modes. An example is the low-frequency peak near 2.81421 cycles day$^{-1}$,
which shows up with an almost identical value in a number of frequency sequences up to 160 cycles day$^{-1}$. These sequences, hereafter referred to as S, are the subject of the present investigation.
We illustrate this in the left panel of Figure~\ref{fig:echelle}, which is an Echelle diagram of the detected frequencies from Paper I, folded with the frequency of 2.81421 cycles day$^{-1}$. Although the diagram examines the frequency peaks in the 6 to 200 cycles day$^{-1}$ range, a few dominant low frequencies were also included in order to match the observed patterns.

\onecolumn
\begin{table}
\caption[]{Frequencies and amplitudes of the $S$ sequences}
\begin{tabular}{rlrrrr rlrrrr}
\multicolumn{2}{c}{Frequency}& \multicolumn{4}{c}{Amplitudes} & \multicolumn{2}{c}{Frequency}& \multicolumn{4}{c}{Amplitudes}\\
cd$^{-1}$ & ID & \multicolumn{4}{c}{parts-per-million} &cd$^{-1}$ & ID & \multicolumn{4}{c}{parts-per-million}\\
& &Q2-Q11&Q2 & Q5/6 & Q7/8  & & &Q2-Q11&Q2 & Q5/6 & Q7/8\\
\hline
\noalign{\smallskip} 
& & $\pm 0.4$ & $\pm 2.0$ & $\pm 0.8$ & $\pm 0.8$ & & & $\pm 0.4$ & $\pm 2.0$ & $\pm 0.8$ & $\pm 0.8$\\
\noalign{\smallskip} 
\multicolumn{5}{l}{Basic sequence} & &  \multicolumn{5}{l}{Combinations with $f_{15}$ (63.3678 cycles day$^{-1}$)}\\
\noalign{\smallskip}
73.2122	&	$S_{25}$	& 3.3	&3.0	&	2.8	&	3.4	& 9.8444	&	$S_{25}$:-$f_{15}$	&&	5.4	&	5.6	&	2.5	\\
76.0223	&	$S_{26}$	& 10.8	&23.1	&	15.5	&	14.0	 & 12.6546	&	$S_{26}$-$f_{15}$	&&	8.6	&	5.0	&	2.5\\
78.8346	&	$S_{27}$	& 4.0	&3.2	&	3.0	&	4.0	& 15.4668	&	$S_{27}$-$f_{15}$	&&	5.7	&	7.1	&	2.6\\
81.6487	&	$S_{28}$	& 11.5	&27.1	&	12.1	&	10.3	& 18.2807	&	$S_{28}$-$f_{15}$	&&	6.8	&	3.6	&	1.7	\\
84.4635	&	$S_{29}$	& 6.6	&1.4	&	6.9	&	5.4	& 21.0957	&	$S_{29}$-$f_{15}$	&&	5.0	&	12.1	&	7.0\\
87.2785	&	$S_{30}$	&10.1	&26.0	&	17.9	&	8.2	& 23.9118	&	$S_{30}$-$f_{15}$	&&	8.4	&	5.1	&	1.5	\\
90.0931	&	$S_{31}$	& 2.7	&8.3	&	3.8	&	1.3	 & 29.5403	&	$S_{32}$-$f_{15}$	&&	9.9	&	3.6	&	1.6	\\
92.9080	&	$S_{32}$	& 6.7	&33.4	&	12.6	&	9.7	& 32.3557	&	$S_{33}$-$f_{15}$	&&	14.1	&	4.8	&	2.8\\
95.7233	&	$S_{33}$	& 2.8	&8.0	&	2.6	&	1.1	& 142.2021	&	$S_{27}$+$f_{15}$	&&	6.7	&	2.3	&	1.8\\
98.5356	&	$S_{34}$	& 4.1	&15.3	&	8.8	&	2.0	& 147.8311	&	$S_{29}$+$f_{15}$	&&	1.5	&	4.2	&	1.7	\\
101.3484 &      $S_{35}$  & 2.0        &4.0 & 2.7 & 1.5 & & & & & & \\
\noalign{\smallskip}
\multicolumn{5}{l}{Combinations with $f_{16}$ (66.2978 cycles day$^{-1}$)} & & \multicolumn{5}{l}{Combinations with $f_{14}$ (60.4341 cycles day$^{-1}$)}	\\
\noalign{\smallskip}
6.9143	&	$S_{25}$:-$f_{16}$	&&	0.9	&	4.1	&	4.3	&  15.5875	&	$S_{26}$-$f_{14}$&&	4.1	&	0.9	&	2.9\\
9.7246	&	$S_{26}$-$f_{16}$&&	17.0	&	5.9	&	5.8	 & 18.4043	&	$S_{27}$-$f_{14}$&&	3.6	&	4.2	&	4.5\\
12.5366	&	$S_{27}$-$f_{16}$&&	5.2	&	5.3	&	5.1	 & 26.8445	&	$S_{30}$-$f_{14}$&&	2.9	&	2.3	&	3.4\\
15.3514	&	$S_{28}$-$f_{16}$	&&	20.1	&	8.7	&	4.0	 & 32.4739	&	$S_{32}$-$f_{14}$&&	5.9	&	3.4	&	1.9\\
18.1659	&	$S_{29}$-$f_{16}$	&&	4.6	&	6.4	&	5.0	& & & & & & \\
20.9814	&	$S_{30}$-$f_{16}$	&&	20.5	&	8.6	&	2.5	&  \multicolumn{5}{l}{Combinations with $f_{13}$ (53.2601 cycles day$^{-1}$)}\\
23.7879	&	$S_{31}$-$f_{16}$	&&	14.2	&	10.2	&	10.6	& & & & & & \\
26.6104	&	$S_{32}$-$f_{16}$	&&	32.5	&	6.9	&	5.5	 & 34.0188	&	$S_{30}$-$f_{13}$	&&	0.0	&	5.8	&	2.7\\
29.4306	&	$S_{33}$-$f_{16}$	&&	13.7	&	20.8	&	17.9	 & 39.6483	&	$S_{32}$-$f_{13}$&&	11.2	&	3.1	&	2.4\\
32.2390	&	$S_{34}$-$f_{16}$&&	16.5	&	6.5	&	3.4	 & & & & & & \\
147.9467	&	$S_{28}$+$f_{16}$	&&	7.3	&	2.2	&	3.1	& \multicolumn{5}{l}{Combinations with $f_{12}$ (50.2782 cycles day$^{-1}$)} \\
159.2061	&	$S_{32}$+$f_{16}$	&&	12.4	&	1.9	&	1.7	&&&&&&\\

&&&&&&	28.5565	&	$S_{27}$-$f_{12}$	&&	2.8	&	3.5	&	5.1	\\
\multicolumn{5}{l}{Related low-frequency peak} &&	34.1854	&	$S_{29}$-$f_{12}$	&&	0.0	&	8.9	&	10.7	\\
2.8142 & $f_1$  & & 117.8 & 18.3 & 12.5  &	39.8156	&	$S_{31}$-$f_{12}$	&&	3.4	&	5.6	&	3.4	\\
&&&&&&	45.4447	&	$S_{33}$-$f_{12}$	&&	1.8	&	3.9	&	3.6	\\
&&&&&&	51.0704	&	$S_{35}$-$f_{12}$	&&	1.3	&	3.0	&	2.2	\\
&&&&&&	129.1130	&	$S_{27}$+$f_{12}$	&&	0.8	&	2.3	&	3.5	\\
&&&&&&	134.7418	&	$S_{29}$+$f_{12}$	&&	0.5	&	3.8	&	3.9	\\
\noalign{\smallskip}

\hline
\end{tabular}
\newline
Identification: $S_{30}$ is the frequency with $m$ = 30 in the formula, $f$ = 2.8519 + $m$*2.81421 cycles day$^{-1}$. The numbering scheme of the frequencies, $f_i$, corresponds to that of Paper I.
\end{table}
\twocolumn														
 
\section{Measured properties of the $S$ sequences}

\subsection{Are the S frequencies exact multiples of a low frequency?}
  
In the data from Quarter 2 of the $Kepler$ mission (hereafter called Q2), a sequence of high frequencies appear to be high multiples of a low frequency at 2.814 cycles day$^{-1}$ with a small zero-point offset. This discovery is made possible by the relatively large amplitudes associated with the low as well as the high frequencies. The additional data from Q5 to Q11 confirm the relationship at lower amplitudes, and provide a considerably increased frequency resolution. This, in turn, confirms
that the high frequencies are not exact multiples of a low frequency, but are equidistantly spaced in frequency with a significant zero-point offset.

Table 1 lists the frequencies identified with the basic S sequence and its combinations with the more dominant (other) pulsation modes.This also shows that the amplitudes vary in a systematic way from quarter to quarter. To achieve the highest precision, we have also calculated a common solution with all the data from Q2 to Q11, covering more than two years. This leads to a high statistical significance for our detections ($\geq$ 4 standard deviations). For the S-sequence combinations with the other, dominant modes, however, we are limited by the small, systematic frequency drifts of the dominant pulsation modes. Since the S sequence produces side lobes to these pulsation modes, the frequencies of the side lobes also change. For these a common solution is not possible; consequently, we only list the separate solutions up to Q7/Q8.

\begin{figure}[htbp]
\centering 
\includegraphics[bb=20 30 840 560,width=\columnwidth, clip]{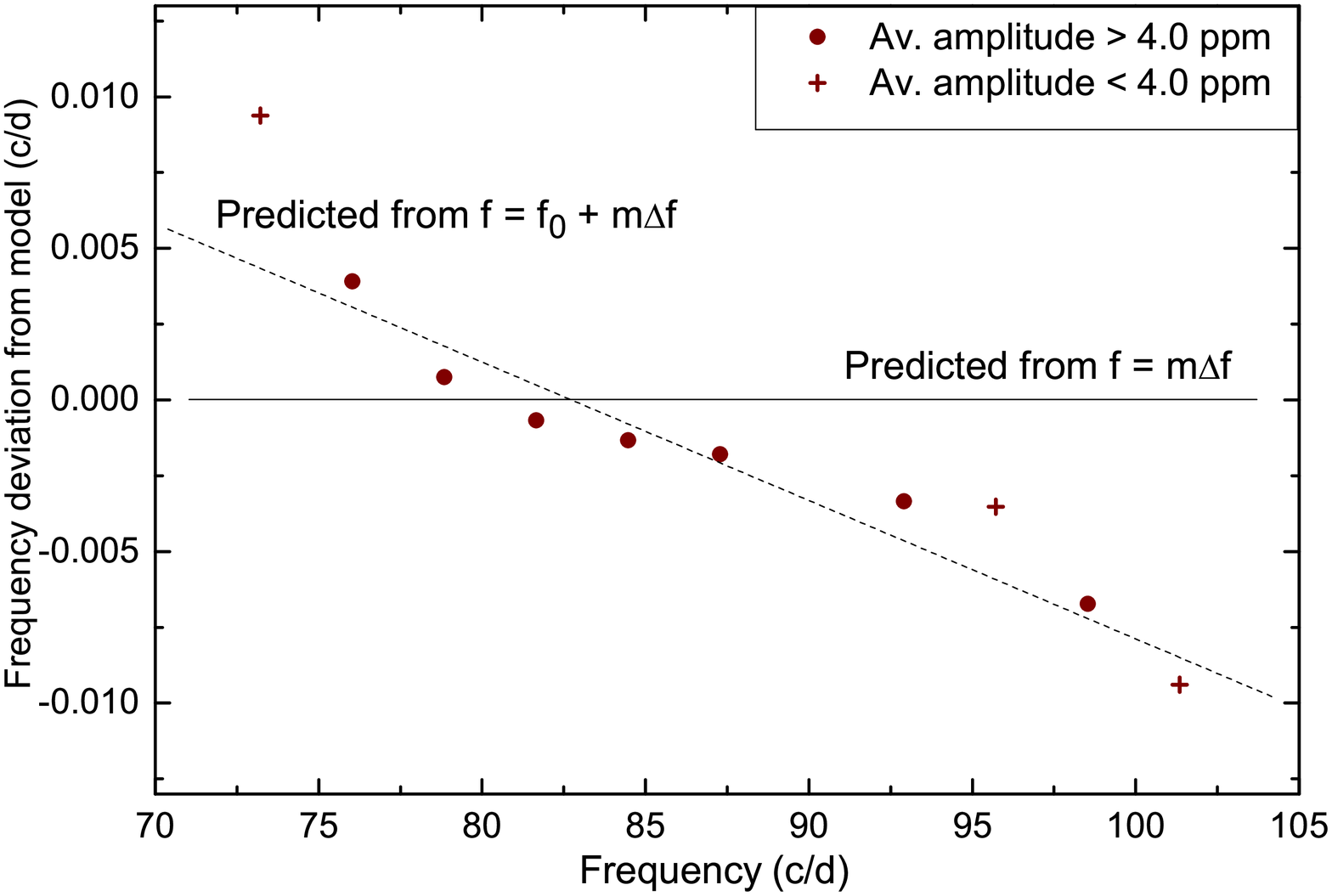}
\caption{Deviations of the S-sequence frequencies from two linear fits with and without an $f_0$ term. The formal error bars are equal to the size of the symbols. The observed frequencies are equidistant with deviations of only 0.001 cycles day$^{-1}$. The spacing corresponds to a dominant low frequency of  2.81421 cycles day$^{-1}$. The diagram also shows that the frequencies have a significant zero-point offset, $f_0$.}\label{fig:2}
\end{figure}

The existence of a significant zero-point offset in frequency is demonstrated in Figure~\ref{fig:2}, where the exact-multiple hypothesis would lead to systematic frequency deviations. We have calculated an optimum least-squares fit from the observed frequencies using the weighting from the observed Q2-Q11 amplitudes. This fit in cycles day$^{-1}$ is:
\begin{equation}
f_m = 2.8519 (\pm 0.0050) + m * 2.81421 (\pm 0.00017),
\end{equation}
where $m$ are integers from 25 to 35. Figure~\ref{fig:2} shows the excellent fit of this equation (dashed line) with an average residual per single point of only 0.0014 cycles day$^{-1}$.

For the pulsation frequency $f_{16}$ at 66.298 cycles day$^{-1}$, the side lobes, displaced by $S$, range from 6.914 to 159.206 cycles day$^{-1}$. The various patterns are illustrated in the right panel of Figure~\ref{fig:echelle}. These agreements and identifications cannot be accidental because of the high frequency resolution ($\Delta f \sim$ 0.0002 cycles day$^{-1}$) of the data set. In fact, we can use the frequencies of these side lobes to check the coefficients derived above. For the side lobes only we obtain
\begin{equation}
f_m = 2.8416 (\pm .0053) + m * 2.81458 (\pm .00018),
\end{equation}
where $m$ once again are integers from 25 to 35. This independent result provides a remarkable agreement with the coefficients derived above from the S sequence.

The offset of 2.8519 cycles day$^{-1}$ vs. a spacing of 2.81421 cycles day$^{-1}$ has important astrophysical implications, which will be explored below. However, we can already reject a simple hypothesis, i.e., that the S sequence is a Fourier series of a non-sinusoidal light-curve shape of $f_1$. In this hypothesis, exact spacing is predicted, as observed. An offset, however, is not possible.

Moreover, the additional $f_0$ term also rules against an interpretation of the S sequence as rotational variability caused by small photometric asphericities on the star.

\subsection{Odd-even amplitude distribution in the frequency sequence, S}

\begin{figure}[htbp]
\includegraphics[bb=10 10 580 730,width=\columnwidth,clip]{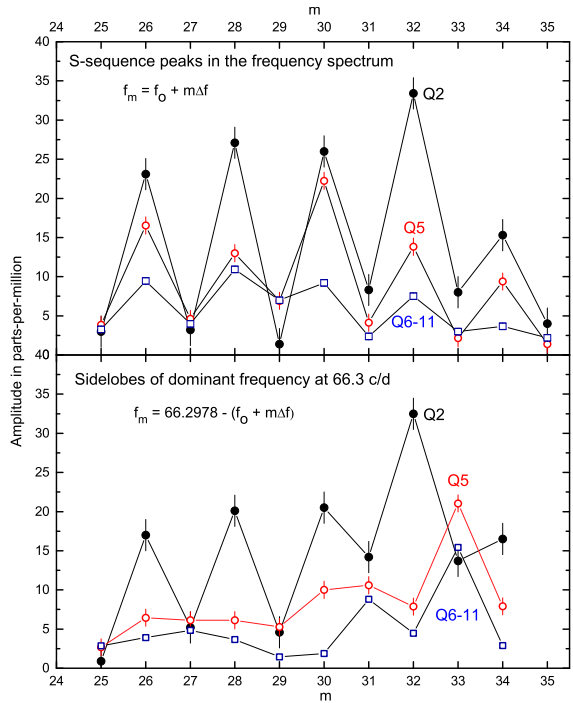} 
\caption{Distribution of amplitudes as a function of the integer $m$ for the S sequence and the side lobes of a dominant frequency at 66.3 cycles day$^{-1}$. From Q2 and Q8,
all amplitudes associated with the S sequences as well as that of the related low frequency at 2.81421 cycles day$^{-1}$, decreased in size. The figure demonstrates the systematic amplitude differences between odd and even values of $m$.}\label{fig:oddeven}
\end{figure}

An important clue to the origin of the sequence is provided by the odd-even sequence of the amplitudes.
In Figure~\ref{fig:oddeven} we show the amplitudes for the different multiples of $m$ for three different time spans. These were selected for the following reasons: Q2 has the largest amplitudes, during Q5 the large amplitudes have already decreased, and from Q6 to Q11 all the peaks are present with small amplitudes, requiring a long data set for an optimum signal-to-noise ratio. The diagram demonstrates an odd-even variation of the amplitudes, with high amplitudes occurring only for even values of $m$. The amplitudes for the odd $m$ peaks are only a few ppm. We note here that despite the small amplitudes the detection of these frequencies is statistically significant due to the low noise of the $Kepler$ data, especially at high frequencies.

\section{Astrophysical interpretations and modeling}

\subsection{Why the high-overtone asymptotic-spacing hypothesis fails}

In this hypothesis, the amplitude differences between odd and even $m$ values can be explained by alternating $\ell$ = 1 and $\ell = 2$ modes and their different amplitudes due to partial light cancellation over the stellar disk. This has already been detected in another $\delta$ Sct star HD 144277 \citep{Zwintz2011}. This would predict a value of the large spacing to be twice the observing frequency spacing, i.e., near 5.6 cycles day$^{-1}$. While such a spacing is within the uncertainties of the known physical parameters of KIC 8054146, present modeling of the observed frequencies shows that the observed frequencies are much too equidistant (as already discussed in Paper I).

\subsection{Why the spot hypothesis fails}

Recent analysis have shown that a large fraction of A/F stars show a rotational modulation of their light output (see \citet{Bal11}, \citet{BBL11}). This asymmetry could be caused by spots. At first sight, the spot hypothesis appears very attractive to explain the S sequence. The reason is that small surface features lead to nonsinusoidal light curves: the equidistantly spaced high frequencies would simply be the higher harmonics of a Fourier series. 

However, there exist serious problems with this hypothesis. The high frequencies are not exact multiples of the observed low frequency. In fact, they are not multiples of {\em any} one or two low frequencies and the measured offset is statistically secure. The frequencies, therefore, are not the result of a Fourier decomposition of an extremely nonsinusoidal light curve. We have confirmed the mathematical argument by attempting to model simple spots: these models failed to give any consistent light curve.

Differential rotation, which may occur at different latitudes, causes two observationally detectable effects. If the (hypothetical) spot has a lifetime of years and is drifting in latitude, then its rotation frequency changes. This is not observed: over two years of observation the frequencies of the S sequence are constant to better than one part in 10$^{5}$.
Furthermore, the observed widths of the Fourier peaks agree with constant frequency values. The lack of additional width or multiple close peaks also argue against multiple spots at slightly different latitudes. We conclude that the S sequence cannot be explained by (simple) spots.

\subsection{Odd-even amplitude differences and high-$m$ prograde sectoral g-modes}

An alternative explanation for the odd-even dichotomy of the S sequence amplitudes can be derived based on the dependence of mode visibility on the surface geometry of the pulsation mode.
It is well known that even-$\ell$ modes have a higher visibility than odd-$\ell$ modes, because odd degree modes suffer from stronger averaging over the visible disc \citep{wad1977,jdd2002}. Consequently, we may interpret the S sequence frequencies as prograde sectoral modes of high spherical degree. Due to the rather high $m$-value, the mode frequencies in the star's system are very low and can be associated to high-order g-modes.

In the case of KIC 8054146, an equator-on view is more likely. With a projected rotation velocity ($v \sin i$) of approximately 300 km s$^{-1}$, the Keplerian break-up rate already amounts to $\approx$0.7 for an equator-on view ($i=90^{\rm o}$). If we decrease the inclination angle, $i$, the star soon reaches the critical rotation rate, which makes a low inclination angle unlikely. A near equator-on view results in a low visibility of modes for which $\ell-|m|$ is an odd number (i.e., north-south asymmetry). In all these cases, there is a node line at the rotation equator and if the star is seen equator-on, the integrated light variation across the surface is diminished by the opposite behavior of the variations of the two sides of node line. The amplitudes fall below the limit of detectability.
An equator-on view in particular favours the visibility of sectoral modes ($\ell$=$|m|$). Due to the mirror effect at zero frequency in the Fourier spectra, both prograde and retrograde modes may populate the same frequency range. However, for rapidly rotating B stars Aprilia et al. (2011) found that retrograde modes suffer a strong damping while prograde sectoral modes are exceptionally weakly damped. Due to the pulsational similarities of A and B stars, the argument can be extended to $\delta$ Sct stars. Consequently, within the family of sectoral modes, we restrict ourselves to prograde modes.

In some sense our case is similar to the case of Rasalhague \citep{monnier2010}, which is also a rapidly rotating A star seen equator-on. In this star sequences of frequencies following the same law
(i.e., $f_0$+$mf_{\rm rot}$) have been observed and were interpreted as high-order g-modes. For Rasalhague $f_0$ is very small ($\approx$0.1-0.3 cycles day$^{-1}$), which leads the authors to the assumption of observing dispersion-free low-frequency modes, which they later specify to be equatorial Kelvin modes. These modes are prograde sectoral modes confined to an equatorial waveguide. 
In the case of KIC 8054146 we chose $f_0=2.8416$ cycles day$^{-1}$ instead of $f_0=0.027$ cycles day$^{-1}$, which would define $m$ to be higher by one integer than in the other case. This, however, would associate the frequencies with highest amplitudes to odd-$m$ modes. A requirement for this interpretation would be an inclination angle of significantly less than 90 degrees, which is ruled out due to the break-up rate as detailed already. 
In addition, there are other important differences between Rasalhague and our case. We do not observe several modes per $m$ value, but only one. Furthermore, our frequencies in the observer's system are significantly higher, with much higher $m$-values in the range of $m$ = 25 - 35.
Consequently, the corresponding surface pattern is characterized by bright and dark sectors.
For example, the angle between two subsequent node lines on the stellar surface amounts to 12$^o$ for a sectoral mode of $|m| = \ell = 29$.

\section{Visibility and excitation of high-degree prograde sectoral modes}

\subsection{Approach}

In the reference system of the star the frequencies of those g-modes causing the S sequence are comparable to the rotation frequency.  To obtain mode properties required for the computation of mode visibility in this low-frequency domain, we adopt the so-called traditional approximation (hereafter: TA) to describe the effect of rotation on oscillation modes.

In the TA, terms proportional to the cosine of latitude ($\theta$) are neglected to allow for the separation of radial and horizontal variables. Instead of spherical harmonics, the latitude-dependence of the eigenfunctions is given by Hough functions, $\Theta$, which are the solutions of Laplace's tidal equations. 

The TA is a standard recipe in geophysics \citep[see][and references therein]{gerkema2008}
 and also gained popularity in studies of low-frequency g-modes in main sequence stars \citep{lee1997,townsend2003a,savonije2005,wad2007}.
We should note, however, that one of the requirements of the TA is that the departure from sphericity is small, and consequently is not valid in rapidly rotating stars such as KIC 8054146. Therefore, only qualitative conclusions can be made with the TA in the given case.

Investigations on the instability of low-frequency g~modes within the TA were performed by \cite{townsend2005a,townsend2005b}, \cite{savonije2005} and \cite{wad2007}. The stability of these so-called slow modes is determined mainly by the separation parameter, $\lambda$, and the angular frequency, $\omega$. Comparisons to methods that avoid the TA were made by \cite{savonije2007} and \cite{lee2006}. \cite{savonije2007} used a 2D approach instead of the TA and finds more prograde modes unstable in comparison with \cite{wad2007}. 
\cite{aprilia2011} examined the effects of mode coupling on the instability in a SPB model and found that, in contrast to zonal and tesseral modes, low-degree prograde sectoral modes in rapidly rotating B stars are almost not affected by damping due to the different dependence of their frequencies on $\Omega$. This also holds true in the case of high-degree modes and A stars.

The visibility of slow modes based on light amplitudes was studied by \cite{townsend2003b}. \cite{wad2007} and  \cite{jdd2007} extended these considerations to RV amplitudes and examined the prospects 
of mode identification. \cite{lee2012} examined the important effect of non-linear mode coupling on the visibility of slow modes.
However, these authors only examined low-degree modes. In this study we examine high-degree modes \citep{balona1999} and neglect mode coupling effects.

Our study is based on codes developed by \cite{wad2007} and \cite{jdd2007}. The codes were slightly modified to allow for the calculation of high-$\ell$ modes. The general approach is as follows: instead of the $\ell(\ell+1)$ term we have a separation parameter $\lambda$, which is determined by solving Laplace's tidal equation for a certain spin parameter, $s=2\Omega/\omega$, azimuthal order, $m$, and taking into account boundary conditions at $\mu=\cos\theta=0$ and $\mu=1$. Our code uses a relaxation technique and, consequently, requires a start value of $\ell$.
In the limit case of $s \rightarrow 0$ (i.e., no rotation), $\lambda \rightarrow \ell(\ell+1)$ and the Hough function is described by a single associated Legendre polynomial except for normalization.
For convenience, we label modes with the $\ell$-value corresponding to $\lambda$(s=0) for the remaining paper.

Linear non-adiabatic calculations yield the complex f-parameter for each mode. This parameter relates the bolometric flux perturbation with the radial displacement and depends on m, $\lambda$, $\omega$ and the stellar model. \textbf{Formally, in the non-adiabatic case, $\lambda$ is a complex quantity (as are the frequency and the spin parameter).
However, in $\delta$~Scuti stars the nonadiabaticity of pulsations is very weak, so that the imaginary part of the frequency is very small relative to the real part. Therefore, in the expression for the spin parameter, $s$,  we use only the adiabatic
(real) part of the frequency and consider $s$ and $\lambda$(s) as real numbers. The same
approach was adopted by Dziembowski et al. (2007): in the adiabatic part of the pulsation code we interpolate the value of $\lambda$(s) from independently computed solutions of the
Laplace tidal equations, and adopt this value for the non-adiabatic part.}
The distortion of the surface is taken into account when computing the limb-darkening of the visible hemisphere as described in \cite{jdd2007}. Our calculation is based on a nonlinear law of \cite{claret2000} as detailed in the appendix of \cite{jdd2007}.
The wavelength-dependent flux derivatives and limb darkening coefficients are computed from a static atmosphere model. 

The visibility of g-modes with intrinsic frequencies close to the rotation frequency was first examined by \cite{townsend2003b}, based on a truncated expansion of $\Theta$ into associated Legendre functions. \cite{wad2007} and \cite{jdd2007} instead performed a two-dimensional integration over the visible hemisphere. The latter approach is also used in our study.

There still are two uncertainties remaining with our chosen approach: the influence of convection on the frequencies of $\delta$ Sct stars and our use of a plane-parallel atmosphere model for such high $\ell$ values. These might restrict us to qualitative, rather than detailed quantitative, conclusions.

\subsection{Application}

\subsubsection{Equilibrium model}

Our subsequent calculations are based on a main sequence model computed with the Warsaw-New Jersey stellar evolutionary code, which was described in \cite{pam1998}. This code evolves one-dimensional star models taking into account uniform rotation by accounting for the effect of the horizontally averaged centrifugal force on the stellar structure.

The examined model has a mass of 1.80 M$_\odot$ and an equatorial rotational velocity of 272 km s$^{-1}$. This rotation rate is less than 2$\sigma$ lower than the measured average $v \sin i$ and its choice will be detailed later. The standard solar composition according to the \cite{asplund09} mixture is adopted as well as rather effective convection ($\alpha_{\rm MLT}=1.8$). Since the oscillations of the given modes are confined to the equatorial region, the effective temperature of the model was chosen to resemble the conditions at the equator, which is slightly cooler than the temperature derived from the integrated disk since it also contains parts of the hotter poles. Table~\ref{tab:model} summarizes the detailed parameters of the equilibrium model.

\begin{table}
  \caption{Parameters of the examined equilibrium model}
  \label{tab:model}
  \centering
  \begin{tabular}{llllll}
     M/M$_\odot$  &  $\log$ $T_{\rm eff}$ & $\log L/L_\odot$ &  $\log g$ &   $f_{\rm rot}$ (c/d)&   $\sigma_{\rm rot}$ \\
\hline
     1.8 & 3.886 & 1.060 & 4.01 & 2.56 & 0.34\\
     \hline
  \end{tabular}
\end{table}

\subsubsection{g-mode properties}

We computed $\lambda$ eigenvalues for different spin parameters for prograde sectoral modes with $m$ = 1--40. The results for $\lambda$ as a function of spin were also checked by comparing them to the asymptotic formulae given by \cite{townsend2003a}, which showed good agreement.

\begin{figure}[htbp]
  
  \begin{center}
    \includegraphics[bb=48 210 544 592,width=\columnwidth]{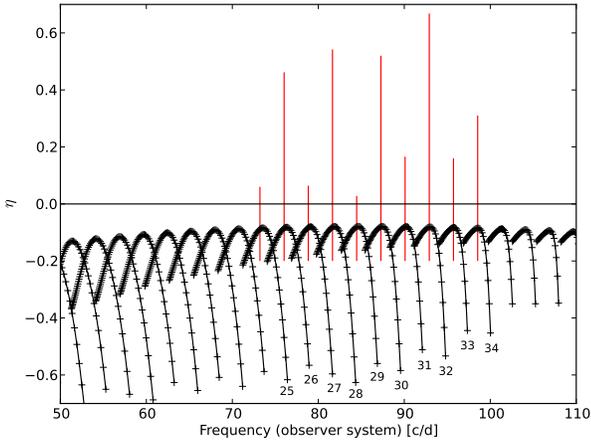}
  \end{center}
    \caption{Normalized growth-rate $\eta$ vs. frequency (in observer's system) for prograde sectoral high-order g-modes (black crosses). Modes with the same $\ell$ are connected by a line and for a subset the corresponding $\ell$ value is denoted. The observed S sequence frequencies are shown with vertical red lines and line lengths proportional to the detected amplitudes.}\label{fig:smopt}
\end{figure}

High-order prograde sectoral g-modes were then computed using a standard oscillation code by replacing $\ell(\ell+1)$ with $\lambda$. 
The results of these computations for the given model are shown in Figure~\ref{fig:smopt}. The normalized growth rate, $\eta=\int_0^R ({\rm d}W/{\rm d}r) {\rm d}r / \int_0^R |{\rm d}W/{\rm d}r| {\rm d}r$, is plotted against the frequency as observed in the inertial system of the observer and compared to the detected S sequence frequencies. Full driving corresponds to $\eta=1$, while full damping corresponds to $\eta=-1$.

It can be seen that for each $\ell$ there is a maximum in $\eta$. The frequency of the maximum mainly depends on the rotation velocity.
\textbf{Consequently, to match the maxima in $\eta$ with the observed 2.814 c/d frequency separation pattern of the S sequence, we adjusted the equatorial rotational velocity to 272 km s$^{-1}$ which corresponds to a rotation frequency of 2.56 c/d (see Figure~\ref{fig:smopt}). This is no disagreement because the observed frequency separations in the S sequence do not correspond to rotational splitting of modes with equal spherical degrees. In addition to the excellent fit of the observed frequency spacing, the even-odd pattern is correctly assigned as well.}

These maxima in normalized growth rate, $\eta$, have a similar nature as the 
 low-frequency maxima in $\eta$ in B stars (which cause SPB type pulsation). In A stars, however, this bump is not as prominent and usually no mode gains sufficient driving. However, as we see in Figure~\ref{fig:smopt} for certain high-$\ell$ modes it is higher than for low degree modes.
The $\eta$-parameter is close to 0.0 at its maximum (for $\alpha$=1.8 and the given mass). Taking into account uncertainties of our simplistic model, both in its global parameters and in description of convection and opacities in  the envelope, it may be possible to obtain marginally unstable modes.

\begin{figure}[htbp]
 
  \begin{center}
    \includegraphics[bb=48 210 544 592,width=\columnwidth]{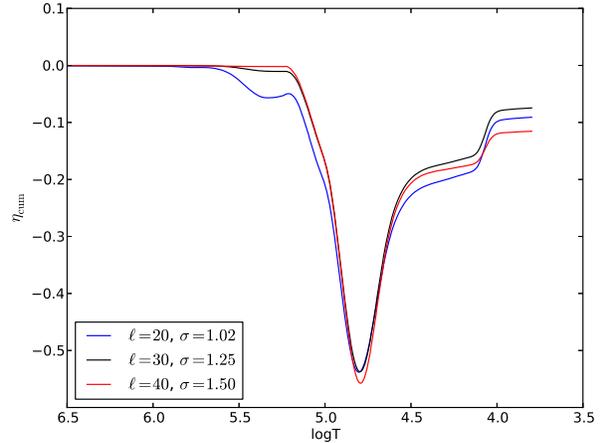}
  \end{center}
  \caption{Cumulative work for three modes at the maximum in $\eta$ but with different $\ell$.} \label{fig:wcum}
\end{figure}

If we consider for each $m$ value only the mode with maximum $\eta$, we also see that we have a local maximum at $m$=30. The reason for this curved shape is illustrated in Figure~\ref{fig:wcum}, which shows the contribution of different stellar layers to driving and damping for the mode with highest $\eta$ for the cases $\ell$=20, 30 and 40. It can be seen that damping near the Z opacity bump (at $\log T \approx $5.3) and differences in the contribution to driving from the \ion{He}{2} and \ion{He}{1}/\ion{H}{0} partial ionization zone (at $\log T \approx $4.65 and 4.1, respectively) are evident for the different modes.

Since this is only a qualitative examination we refrain from listing the exact mode properties in a table. Generally, the parameters of the modes which correspond to the observed counterparts are as follows: the spin parameter ranges from s=0.55 (at $m$=25) to s=0.44 (at $m$=35) and \textbf{the nondimensional frequency, $\sigma=\sqrt{4 \pi G <\rho>} \omega$, from 1.24 to 1.53. Moreover, the radial orders span the range from n=54 (at $m$=25) to n=60 (at $m$=35).}

A potential problem remains:  for each $\ell$, we observe only one frequency, but several modes of subsequent radial order are predicted. Why is only one mode excited? Formally, all these modes are stable, but due to $\eta$ being close to 0, only a minor contribution to driving is sufficient to excite these modes. Consequently, it may be possible that with increased driving only the modes at maxima become marginally unstable, forming the given equidistant pattern.

\subsubsection{Mode visibility}

\begin{figure}[htbp]
  
  \begin{center}
    \includegraphics[bb=48 210 544 592,width=\columnwidth]{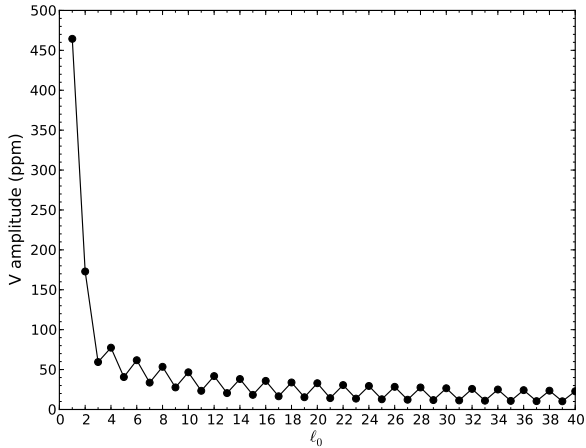}
  \end{center}
    \caption{Mode visibility of prograde sectoral high order g-modes in the Geneva V band.}\label{fig:visibility}
\end{figure}

We computed the visibility for the subset of modes, which is located at maxima in $\eta$. Due to lack of atmosphere data for the Kepler band we computed the results for the Geneva $V$ band. The results are shown in Figure~\ref{fig:visibility}.

To obtain similar amplitudes as observed, the intrinsic mode amplitude, $|\varepsilon|$, was adjusted to 
0.0001, which implies rather small intrinsic amplitudes.
In the range between $m$=25 and 35, the ratio between amplitudes of even-$\ell$ to odd-$\ell$ modes is $\approx$2.5, which is similar to the observations.
Moreover, there is only a small general decrease of the given alternating pattern towards higher $\ell$ values, significantly increasing the probability of detecting high-degree modes.

\section{Conclusion}

In this paper, we have examined a number of equidistantly spaced frequency series, S, in KIC 8054146 with a spacing of 2.814 cycles day$^{-1}$. These are interpreted in terms of prograde sectoral modes as well as a small amplitude modulation of the dominant pressure modes. The basic sequence is found to be multiples of 2.814 cycles day$^{-1}$ (up to 35) with a significant zero-point offset. The spacing and the existance of an offset are similar to the behavior of the reported prograde sectoral mode patterns detected in Rasalhague by \cite{monnier2010}, albeit with smaller multiples. We note again that the equidistant spacing is caused by the observer's frame of reference. 

For KIC 8054146, the measured projected rotational velocity of 300 km s$^{-1}$ and the corresponding fraction of the Keplerian break-up velocity of 0.7 indicate a higher probability of a near equator-on view.
In such a geometrical configuration modes with north-south symmetry, in particular sectoral modes, have the highest chances to be detected. As detailed in Section 4.3, prograde modes are the most viable candidates among sectoral modes.

The observed odd-even dichotomy of the amplitudes of the S-sequence frequencies can be explained by associating these frequencies to odd and even azimuthal order respectively.
We computed the visibility of prograde sectoral modes with a formalism based on the TA. Assuming a fixed intrinsic amplitude for all modes, we find that the computed amplitudes of prograde sectoral g-modes do not decrease quickly in an equator-on view. Hence, it becomes possible to detect them.
The frequencies and amplitudes of the S sequence can be approximately fitted with a 1.8 solar mass model, which rotates with an equatorial rotation velocity of approximately 272 km s$^{-1}$. 

A remaining uncertainty with our interpretation is that the modes with the highest normalized growth-rates are still predicted to be stable. However, the possible instability of these modes is within the uncertainties of the given stellar equilibrium model in terms of opacities, chemical composition, efficiency of envelope convection or stellar mixing processes in general, and also interactions between pulsation and rotation. Generally, if our hypothesis is correct, it would require marginal instability only for the modes at maxima in the normalized growth rate, i.e., with a specific radial order, while the other modes surrounding this maximum are stable. This is a very sharp criterion and hence has a low probability of being observed. Consequently, another selection mechanism might also be involved. 
Although for KIC 8054146 some open questions remain, our qualitative theoretical study shows evidence that high-degree prograde sectoral modes may be driven and observed in $\delta$~Scuti stars.

\acknowledgments

It is a pleasure to thank Richard Townsend, Wojtek Dziembowski, and Luis Balona for helpful discussions. This investigation has been supported
by the Austrian Fonds zur F\"{o}rderung der wissenschaftlichen Forschung through project P\/21830-N16. 
PL and AAP acknowledge partial financial support from the Polish NCN grants 2011/01/B/ST9/05448 and 2011/01/M/ST9/05914.
The authors also wish to thank the {\it Kepler} team for their outstanding efforts which have
made these results possible.  Funding for the {\it Kepler} mission is provided by NASA's Science Mission Directorate.

\end{document}